\documentclass[11pt,a4paper,titlepage]{article}
\usepackage{amsfonts}
\usepackage{amssymb}
\usepackage{amsthm}
\usepackage{dsfont}
\usepackage{amsmath}
\usepackage{latexsym}
\usepackage{bbm}
\usepackage{stmaryrd}
\usepackage{graphicx}
\usepackage{caption}
\usepackage{authblk}
\usepackage{booktabs}
\usepackage{tabularx}
\usepackage{threeparttable}
\newcolumntype{L}{>{\raggedright\arraybackslash}X}
\newcolumntype{C}{>{\centering\arraybackslash}p{1.3cm}}
\usepackage[font={small,sf},labelfont=bf]{caption}
\usepackage[justification=raggedright,singlelinecheck=false]{caption}
\usepackage[
backend=biber,
style=numeric,
sorting=none,
url=false,
doi=true,
eprint=true,
maxbibnames=3,   
maxcitenames=2   
]{biblatex}
\AtEveryBibitem{\clearfield{note}}
\begin{document}

\begin{titlepage}
	\title{\textbf{Revisiting the Fermi Paradox: \\ 
			Does the Math Hold up?}}
	\author{Philipp N. Spahn\footnote{Independent Researcher. San Francisco, CA, USA} \& Elmar Teufl\footnote{Department of Mathematics, University of Tübingen, Germany}}
	\date{\today}
	\maketitle
\end{titlepage}

\tableofcontents
\newpage

\section{Abstract}
In this paper, we present a stochastic model to address the famous Fermi Paradox, i.e. the supposed contradiction between the abundance of intelligent life in the universe and the absence of any evidence of it. The model assumes that life can arise and evolve on habitable planets with a certain chance, and that we can detect life on other planets within a certain distance. The parameters of the model include the rates, at which life arises, evolves, vanishes, or advances to ``Space Age", i.e. a stage where it gains the capability of interstellar communication. Given any numerical choice of these parameters, the model estimates a Contact Probability, i.e. the probability of having received a signal from an extraterrestrial civilization during the ongoing Space Age on Earth. Based on a critical probability threshold, the model can identify regions in the parameter space where Contact Probability is low, i.e. where the absence of contact is the expected outcome rather than a paradox. Monte-Carlo exploration of the parameter space shows that, even when assuming a high frequency of habitable planets, the median Contact Probability across a wide range of parameter choices is under 0.42, and under 0.06 if we assume that habitable planets generally follow an ``Earth-like" trajectory, with life requiring roughly 4.5 billion years to advance to Space Age. Thus, if Earth was a typical planet with regard to its evolutionary timeline, the Fermi Paradox would likely not be a paradox.
\newpage

\section{Introduction}
The Fermi Paradox \cite{hart_explanation_1975, sagan_direct_1963} has been a conundrum for over half a century. It is generally thought to have originated from a casual lunch conversation at the Los Alamos National Laboratory in 1950 when Enrio Fermi famously wondered ``... but where is everybody?" referring to the observed absence of extraterrestrial visitors on Earth despite their expected existence in the vastness of the galaxy. A plethora of ideas have been brought forward to explain how this apparent paradox might be avoided \cite{webb_seventy-five_2015}. Still, the Fermi Paradox remains to be largely seen as an unresolved mystery, both across the scientific community and in popular culture. While at first glance the expectation of life outside the solar system may seem preposterous, or at best nothing more than a fancy thought experiment entertained by an unworldly intellectual elite, modern astronomy has found irrefutable evidence of planets outside the solar system that could support life \cite{bohl_probing_2026}. In fact, life appears to have arisen surprisingly quickly on Earth, with the oldest microbial fossils thought to have appeared only a few hundred million years after the formation of our planet. Thus, even a skeptic would have to acknowledge that despite the probability of life to arise being possibly very small, and its evolution to intelligence even smaller, the sheer number of habitable exoplanets could accumulate to a reasonably high number and make the emergence of another Space Age civilization possible, or even likely. Yet, despite decades of scanning the sky for suspicious transmissions, the search has---apart from few cases of enigmatic, yet non-reproducible anomalies \cite{mendez_arecibo_2025}---found no evidence of extraterrestrial signals. While it is being debated whether Fermi himself thought of this contradiction as a paradox \cite{gray_fermi_2015}, or whether he would even have had enough reason to do so, given that Deep Space exploration was still in its infancy in 1950, the putative paradox remains a popular topic and is today typically phrased more specifically: ``Given the large number of habitable exoplanets, which should result in the existence of Space Age civilizations, why has there not been any evidence of them?" The classic quantitative tool to address the Fermi Paradox is the Drake formula \cite{drake_radio_1965}, which enumerates how many Space Age civilizations should, on average, exist at any point in time in our galaxy. The formula contains a total of seven parameters, only three of which are subject to observation and thus not entirely up for speculation. Hence, the Drake formula can yield an almost arbitrary (non-negative) number of Space Age civilizations depending on the specific assumptions. In addition, a number of limitations of the formula have been pointed out: First and foremost, it has been noted that the total number of Space Age civilizations in the entire galaxy is likely an irrelevant metric with regard to the Fermi Paradox, since their distances to Earth are not taken into account. Thus, even if there was a large number of co-existing Space Age civilizations, they may never make contact with one another, given the constraints in detection methods over large distances \cite{smith_broadcasting_2009}. Furthermore, the Drake formula provides only a steady-state average, while the number of Space Age civilizations would be more naturally described as a function of time \cite{glade_stochastic_2012,devito_fermi_2024}. Finally, life on Earth entered its Space Age only relatively recently, which needs to be considered when interpreting the lack of extraterrestrial signals \cite{grimaldi_inferring_2023}. The non-detection of signals over a listening period of only 60 years obviously poses a weaker Fermi Paradox than a listening period of 6,000 years would. Therefore, even if the parameters in the Drake formula were known, the Fermi Paradox would not be resolved.\\ 
\newline Here we propose a new angle to address the Fermi Paradox by describing the evolution of life on exoplanets as a Markov Chain stochastic process, and by incorporating the detection distance and Space Age duration into the model. Assuming a uniform distribution of exoplanets in our local galactic neighborhood, which undergo evolution independently from each other, the model returns a Contact Probability, being the probability that emissions from an exoplanet would have arrived during the ongoing Space Age on Earth. Monte-Carlo simulation of the parameters in our model indicates that Contact Probabilities tend to be low, even under rather generous assumptions, which sheds new light on solutions to the Fermi Paradox from a mathematical point of view.

\newpage

\section{Model Description}
We lay out a mathematical description of the Markov model below. Refer to section \ref{Section:ModelSummary} and Fig.~\ref{Fig:ModelSummary} for a non-mathematical summary of the model, which is sufficient to follow the rest of the article.

\subsection{Life Evolution as a Markov Chain}
We follow a common definition of ``life" in astrobiology as a ``self-sustaining chemical system capable of Darwinian evolution". For our model, we consider a planet ``habitable" if it presents conditions that enable life to emerge and to evolve into a complex enough form to engage in interstellar communication. We note that this does not imply that every such star hosts a planet with life, since the time required to spawn life (which will be another model parameter) may exceed the lifespan of the star. We model evolution on a habitable planet with a Markov process $X=(X_t)_{t\ge0}$ with four different stages $\{0,1,2,3\}$ and transition matrix family $\Pi=(\Pi_t)_{t\ge0}$, with $t$ in years and $t=0$ being the time of the formation of the planet (which we equate with the formation of the star for simplicity). We define the four stages as:
\begin{itemize}
	\item 0: Planet has no life
	\item 1: Planet has primitive life
	\item 2: Planet has intelligent life
	\item 3: Planet has life in Space Age
\end{itemize}
We define ``Space Age" as an era, during which life is both detectable and capable of detecting signals across interstellar distances (with the numerical value of this distance being another model parameter). We define ``intelligent life" broadly as an era, during which life develops technology, and ``primitive life" as all remaining forms of life. We follow the Drake formula in distinguishing only these three stages of life and abstain from including additional stages life could evolve through (e.g. \cite{roudenko_exploring_2026}) in order to minimize model assumptions.
\\

\begin{figure}[htbp]
	\centering
	\includegraphics[width=10cm]{"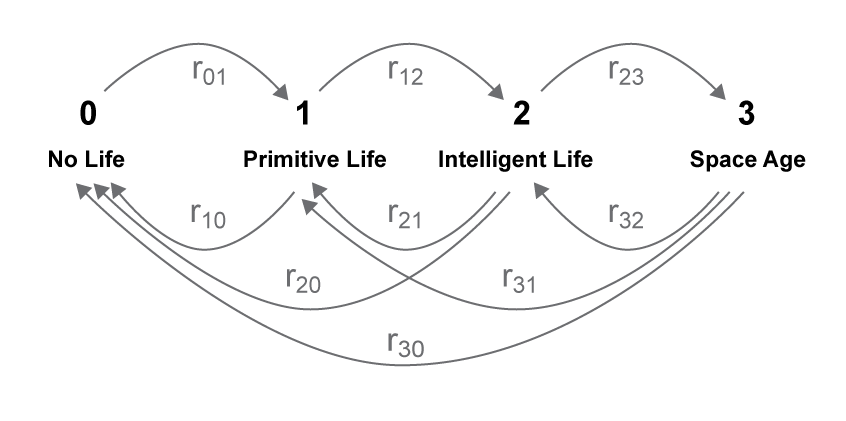"}
	\caption{Evolution is modeled with a Markov process that transitions through four stages with the indicated rates.} \label{Fig:JumpChain}
\end{figure}

The process jumps between these stages with rates $r_{ij}$ (Fig.~\ref{Fig:JumpChain}). Let $R = (r_{ij})_{0\leq i,j \leq 3}$ be its jump matrix, i.e. $r_{ij}$ is the probability of jumping from stage $i$ to stage $j$, provided the process does leave stage $i$. We assume life does not skip stages when moving to a more advanced stage, i.e. $X_0=0$, and $r_{02}=r_{03}=r_{13} = 0$.
\begin{equation*}
	R = 
	\begin{bmatrix}
	0 & 1 & 0 & 0 \\
	r_{10} & 0 & r_{12} & 0 \\
	r_{20} & r_{21} & 0 & r_{23} \\
	r_{30} & r_{31} & r_{32} & 0
	\end{bmatrix}
\end{equation*}
We review the basic properties of Markov Chains needed in this paper in the Mathematical Appendix, Section \ref{Section:basicmarkov}.

\subsection{Model Refinements}
We want to make two adjustments to the basic model above.

\subsubsection{Gradual Extinction}
We first want to simplify the jump matrix $R$ to reduce the number of free parameters and eliminate the possibilities of a complete eradication of all life from an advanced stage, and instead let extinctions always transition through a ``Primitive Life" stage first (however short this may last). While this certainly makes the model less general, we believe this modification is reasonable, based on observations from Earth's evolutionary history where even major extinction events occurred gradually and affected the more complex lifeforms first. Hence, we set 
\begin{equation*}
	r_{20} = r_{30} = 0
\end{equation*}
With that, the jump matrix simplifies to:
\begin{equation*}
	R = 
	\begin{bmatrix}
		0 & 1 & 0 & 0 \\
		\alpha & 0 & 1-\alpha & 0 \\
		0 & \beta & 0 & 1-\beta \\
		0 & \gamma & 1-\gamma & 0
	\end{bmatrix}
\end{equation*}
with $\alpha, \beta$ and $\gamma$ being numbers between 0 and 1, describing the chance of life to reset to a less advanced stage. 

\subsubsection{Red Dwarfs}\label{Section:ModelModRedDwarf}
Stars vary in mass, temperature, and brightness, and can be classified into seven main types (\textit{O,B,A,F,G,K,M}). While all star types have habitable zones, i.e. distance zones in which planets can be habitable, the conditions in each may depend on the type of star. While we want to avoid implementing assumptions into our model based on speculations about the specific conditions around each type of star, we do want to account for the violent nature of Red Dwarfs (K- and M-type stars) in our model, since conditions around these stars are generally considered to be major constraints for the evolution of complex life \cite{shields_habitability_2016}.\\
We thus want to incorporate the increased extinction risk around Red Dwarf by modifying the life extinction rate $\alpha$ with a \textbf{Red Dwarf Ferocity} factor $M\geq 1$ and $K\geq 1$ for planets around M- and K-type stars, respectively. Thus, for planets around these types of stars we have the jump matrix:
\begin{equation*}
	R_M = 
	\begin{bmatrix}
		0 & 1 & 0 & 0 \\
		M\cdot\alpha\cdot\rho^{-1} & 0 & (1-\alpha)\cdot\rho^{-1} & 0 \\
		0 & \beta & 0 & 1-\beta \\ 
		0& \gamma & 1-\gamma & 0
	\end{bmatrix}
\end{equation*}
with $\rho:=M\cdot\alpha + 1 - \alpha$ being the row sum for normalization purposes (all rows need to sum up to 1), and analogously for K-type stars.

\subsection{Space Age on Earth and Detection Radius} \label{Section:DetectionRadius}
We want to use our model to address the question of how probable it is for our civilization to learn of the existence of another civilization if the galaxy contained civilizations in Space Age, i.e. civilizations that (i) are emitting signals into Space and (ii) are able to receive signals from Space, which both of these activities being continuous in time. We do not specify the nature of these signals, but want to assume that all civilizations in our model would be able to send and interpret the same signals, e.g. radio waves. For signal emission, we do not distinguish intentional signals and unintentional signals, for example leakage of radio transmission into Space from commercial or military radio sources. Likewise, for signal reception, we do not distinguish intentional ``listening" efforts (such as the SETI program) and serendipitous signal receptions, e.g. through radio telescopes, which are scanning the sky for other purposes, but would detect extraterrestrial signals if they occurred. We assume that each civilization would emit signs of its existence in all directions once it reached Space Age, and likewise receive signals from all directions. Thus, in our model, signals are omnidirectional point sources, akin to beacons, which last throughout the duration of Space Age. \\

With this broad definition, we observe that our civilization (i) has been sending radio signals into Space since roughly the 1940s in a nearly omnidirectional fashion, and (ii) has operated large radio telescope arrays enabling nearly continuous and omnidirectional ability to detect radio signals from Space since the 1960s. We thus consider life on Earth to meet the definition of being in Space Age, and assume that Space Age on Earth has been continuous for a total of approximately \textbf{65 years} so far.  \\

We note that the above definition of Space Age implies a radius, within which a civilization would be able to receive interstellar signals. This \textbf{Detection Radius} would be determined by the strength of the emitted signals and the sensitivity of the receiving equipment. As any energy emitted from a point source declines with the square of the distance traveled, the signal range would be rather short, a few light years at best, for the kind of unintentional signals emitted from Earth, i.e. low energy radio noise. But it is certainly conceivable that extraterrestrial civilizations emit higher-energy signals. For intentional signals, such as high-energy radio beams, the range would be much larger, easily in the thousands of lightyears. But sending such narrow-band, high-powered beams at every exoplanet in an omnidirectional and continuous fashion would require an elusive amount of energy. Possibly, Earth could receive a targeted signal because our biosignature may have been previously spotted by the sender. However, based on our current understanding of detection methods---such as planetary transits---a detection of Earth from all directions would be implausible, and any practical range would likely be confined to tens of lightyears. Advanced civilizations might possess techniques beyond our own, but even superior technology would still face physical sensitivity limits. Consequently, for the purpose of this model, we want to subsume all the above mentioned types of signals (unintentional leakage and intentional messages) into a single entity and assume its range to be limited by a Detection Radius. To leave room for superior technology, a Detection Radius on the order of hundreds of lightyears seems like a reasonably generous assumption. Similar concepts of a critical communication distance were adopted in e.g. \cite{smith_broadcasting_2009,prantzos_probabilistic_2020,grimaldi_signal_2017,balbi_longevity_2021}.

\subsection{Frequency, Age, and Distribution of Exoplanets} \label{Section:StarAgeAndPos}	

\begin{figure}[htbp] 
	\centering
	\includegraphics[width=15.5cm]{"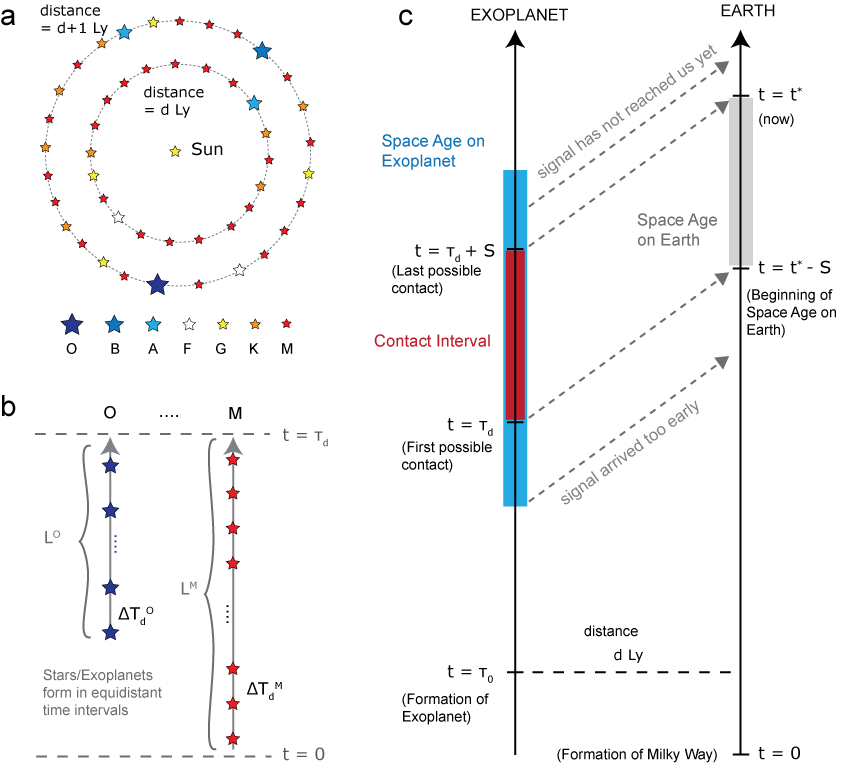"}
	\caption{\textbf{a:} The habitable stellar neighborhood is modeled as concentric and equidistant shells around the sun. \textbf{b:} In each shell, the stars present at a certain time $t=\tau_d$ are assumed to span all ages, i.e. the youngest star has just recently formed and the oldest star has reached the maximum lifespan $L$, with the remainder having formed at discrete time intervals in between. \textbf{c:} In order for Earth to receive a signal from an exoplanet, Space Age on the exoplanet needs to (at least partially) overlap with Space Age on Earth, taking into account the signal's travel time.} 
	\label{Fig:StarFormContactInt}
\end{figure}

The number and position of stars in our neighborhood is fairly well understood. Based on a commonly used local density estimate of 0.14 stars per cubic parsec, we assume there are roughly 
\begin{equation*}
U(d) = 0.14\cdot\frac{4}{3}\pi\cdot \left(\frac{d}{3.26}\right)^3
\end{equation*}
stars in a sphere with radius $d$ lightyears around the sun. For simplicity, we will assume that these stars are uniformly distributed in space, i.e. located on concentric shells around the sun, with a fixed distance of 1 lightyear between these shells (Fig.~\ref{Fig:StarFormContactInt}a). Thus, the number of stars on each shell can be estimated as:
\begin{equation*}
	u(d):=U(d)-U(d-1)=0.14\cdot\frac{4}{3}\pi\cdot3.26^{-3}(3d^2-3d+1)\  (d\in\mathbb{N}).
\end{equation*}
We will also assume that at each distance the total number of stars $u(d)$ is a sum of eight different star types: 
\begin{equation*}
	u(d) = u_O(d) + u_B(d) + u_A(d) + u_F(d) + u_G(d) + u_K(d) + u_M(d)
\end{equation*}
with the number of each star type given by their average abundance (Table \ref{Tab:StarTypes}), e.g. $u_F(d) = 0.03\cdot u(d)$. We assume that a percentage $h$ of stars harbors a habitable planet, i.e. we have $N_d:=h\cdot u(d)$ habitable planets in distance $d$ lightyears. While theoretically habitable zones can be maintained around dying and dead stars, we only consider main sequence stars in this model, and exclude Brown Dwarfs. For simplicity, we also omit the possibility that stars could can more than one habitable planet. \\
\begin{table*}[h]
	\caption{Star types and their average abundances and main sequence lifespans considered in the model.}
	\label{Tab:StarTypes}
	\begin{tabularx}{0.85\linewidth}{L *{7}{c}}
		\toprule
		& O & B & A & F & G & K & M \\
		\midrule
		Abundance [\%] & 0.01 & 0.13 & 0.60 & 3.00 & 7.60 & 12.10 & 76.56 \\
		Lifespan [$10^6$ years] & 4 & 15 & 800 & 4500 & 12000 & 25000 & 700000 \\
		\bottomrule
	\end{tabularx}
\end{table*}

Star age is inherently difficult to measure as it relies on proxy measurements like rotation speed, velocity of proper motion, or magnetic activity. While it is known that young stars are mostly located in star-forming regions, we assume that stars are uniformly distributed in age in our local neighborhood. For the purpose of our model, we therefore want to let the stars at every distance span in age from 0 to their respective average lifespan $L$ (Table \ref{Tab:StarTypes}). For example, on the shell at distance $d$ lightyears, where we have a total of $N_d$ habitable planets, we assume that these planets (and their hosting stars) formed in discrete time intervals in the past:  
\begin{equation}\label{Eqn:DeltaT}
	\Delta T_d := \frac{L}{N_d}
\end{equation}
with $L$ (and thus $\Delta T_d$) depending on star type (Table \ref{Tab:StarTypes}) (we round $\Delta T_d$ to a natural number) (Fig.~\ref{Fig:StarFormContactInt}b). Since for K-, and M-type stars, the average lifespan extends the estimated age of the Milky Way, we use $L = 10$ billion years in these cases.

\subsection{Contact Probability and Contact Interval} \label{Section:ContactProb_Interval}
With each habitable exoplanet in our local neighborhood going through life evolution described by our Markov model above, we are interested in the \textbf{Contact Probability $p$}, i.e. the probability of having received a signal from an exoplanet within the ongoing Space Age on Earth. We assume that life evolution on separate habitable planets proceeds independently from each other, follows the same process $X$ with the same jump rates and stage durations, and starts at the time of formation of the planet. That is, every habitable planet follows the process $X=(X_t)_{t\ge\tau_0}$ with $X_{\tau_0}=0$ and $\tau_0$ being the time of formation of the planet. \\

We first consider a single exoplanet in distance $d$ from Earth (in lightyears), which formed at a time $t=\tau_0$. We assume that the planet will emit signals as soon as its civilization enters Space Age (i.e. as soon as $X_t=3$). In order for us to receive a signal, Earth needs to be in Space Age as well. More precisely, the planet needs to have been in its Space Age at a time that aligns with Space Age on Earth, given the time needed for the signal to bridge the distance between them (Fig.~\ref{Fig:StarFormContactInt}c) (A similar approach was used e.g. in \cite{balbi_impact_2018}):\\
If we let $S=65$ years be the duration of the ongoing Space Age on Earth, and $t=t^*$ be now, then the earliest point in time where we could have received a signal would be $t=t^*-S$. Thus, in order to receive a signal from an exoplanet during Space Age on Earth, i.e. the time interval $[t^*-S,t^*]$, the exoplanet needs to have life in Space Age at least for some time during the interval 
\begin{equation*}
	I_d := [\tau_d,\ \tau_d+S] := [t^*-S-d/c,\ \ t^*-d/c]
\end{equation*}
with $c$ being the speed of light. We call this interval the \textbf{Contact Interval} for this particular exoplanet. Thus, the Contact Interval of an exoplanet is the only time interval, during which a signal from this particular planet could have reached Earth during our Space Age. With this, the event
\begin{equation*}
	C:= \textit{"Earth receives a signal during Earth's ongoing Space Age"}
\end{equation*}
can now be rephrased as \textit{"At least one of the N exoplanets within Earth's Detection Radius was in Space Age at some point during its Contact Interval"}. If we set
\begin{equation*}
	C_n := \textit{"Planet n is in Space Age for some time during its Contact Interval"}
\end{equation*}
then the Contact Probability is
\begin{equation*}
	p:=\text{P}(C)=\text{P}\left(\bigcup_{n=1}^{N} C_n\right)
\end{equation*}
As the probability that any event happens is equal to 1 minus the probability that none of the events happen, we can express this as
\begin{equation*}
	p = 1-\text{P}\left(\bigcap_{n=1}^{N} \overline{C_n}\right) =1-\prod_{n=1}^N \text{P}(\overline{C_n})=1-\prod_{n=1}^N [1-\text{P}(C_n)] 
\end{equation*}
since the $C_n$ are independent events. And since we can group the $N$ planets by distance, each group containing $N_d$ habitable planets, we have 
\begin{equation}\label{Eqn:pC}
	p = 1-\prod_{d=1}^D\prod_{n=1}^{N_d} [1-\text{P}(C_n)]
\end{equation}
with $D$ being the Detection Radius. \\
Thus, in order to compute the Contact Probability, we need to compute $P(C_n)$: the probability for a given exoplanet to be in Space Age at least for some time during its Contact Interval. We  present the calculation of this probability in the Mathematical Appendix, Section \ref{Section:P(Cn)}.

Throughout the rest of this paper we want to compute (or estimate) the Contact Probability for a number of example scenarios, and analyze more broadly how the Contact Probability changes for different choices of our model parameters (Table~\ref{Tab:modelparms}).

\subsection{Model Summary} \label{Section:ModelSummary}
\begin{figure}[htbp] 
	\centering
	\includegraphics[width=15.5cm]{"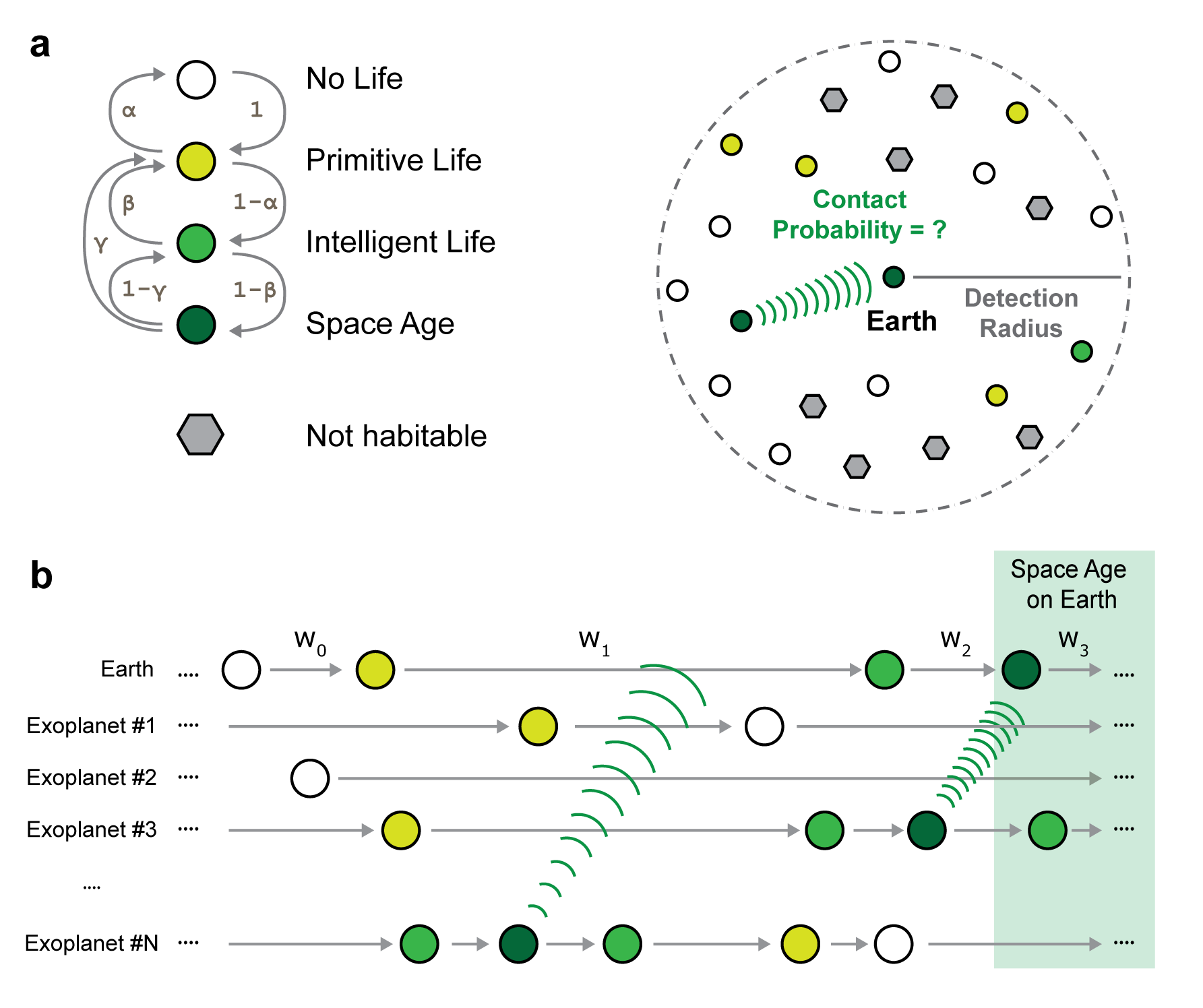"}
	\caption{\textbf{a:} Life on habitable planets evolves through four stages, and transitions between these stages with the indicated rates $\alpha, \beta$, and $\gamma$. A planet within the Detection Radius that has reached Space Age will emit signals which can be received on Earth. The Contact Probability describes the probability for such a signal reception to have occurred since the beginning of Space Age on Earth. \textbf{b:} Life evolves independently on different habitable exoplanets. The times life spends in each stage ($w_0, w_1, w_2$, and $w_3$) are stochastic variables and can differ between planets. In the hypothetical example, Exoplanet \#1 evolved primitive life, which subsequently died out. Exoplanet \#2 has not developed any life yet. Exoplanet \#3 reached Space Age and sent signals, which were received on Earth once it also reached Space Age. By the time the signal did reach Earth, Space Age had already ended on Exoplanet \#3. Exoplanet \#N reached Space Age at a time when only primitive life was present on Earth. By the time Earth reached its Space Age, life on the planet had already died out.} 
	\label{Fig:ModelSummary}
\end{figure}

In our model, life on habitable planets evolves through four stages (``No Life" / ``Primitive Life" / ``Intelligent Life" / ``Space Age") with Space Age being the stage, during which life is both detectable on other planets and itself able to detect life on other planets (within a certain Detection Radius). Life advances (or reverts) across these stages stochastically (Fig.~\ref{Fig:ModelSummary}a), and independently on different planets. In order to make contact with extraterrestrial life within the Detection Radius, Space Age on Earth and on the other planet need to be aligned, taking into account the delay needed for signal transmission (Fig.~\ref{Fig:ModelSummary}b). The probability of such an event to have happened within the ongoing Space Age on Earth is called the \textbf{Contact Probability}. Space Age is assumed to have been ongoing on Earth for ca. 65 years. The model has a total of 10 parameters (Table \ref{Tab:modelparms}).\\

\begin{table*}[h]
\caption{Model parameters}
\label{Tab:modelparms}
\begin{threeparttable}
	\renewcommand{\arraystretch}{1.2}
	\begin{tabularx}{\linewidth}{p{5cm} c L}
		\toprule
		Parameter & Symbol & Description \\
		\midrule
		No Life Duration & $w_0$ & Average time a planet remains in the ``No Life" stage \\
		Primitive Life Duration & $w_1$ & Average time a planet remains in the ``Primitive Life" stage \\
		Intelligent Life Duration & $w_2$ & Average time a planet remains in the ``Intelligent Life" stage \\
		Space Age Duration & $w_3$ & Average time a planet remains in the ``Space Age" stage \\
		Life Extinction Rate & $\alpha$ & Probability for all life to go extinct, i.e. for a planet to revert from ``Primitive Life`` to ``No Life'' \\
		Intelligence Extinction Rate & $\beta$ & Probability for intelligent life to go extinct, i.e. for a planet to revert from ``Intelligent Life`` to ``Primitive Life'' \\		
		Space Age Extinction Rate & $\gamma$ & Probability for Space Age to trigger the extinction of intelligent life, i.e. for a planet to revert from ``Space Age`` to ``Primitive Life'' \\		
		Red Dwarf Ferocity & $M$ & Fold-increase multiplier in Life Extinction Rate on planets around M-type Red Dwarfs \tnote{a} \\
		Habitability & $h$ &  Percentage of stars harboring a habitable planet \\
		Detection Radius & $D$ &  Maximum distance (in lightyears), across which planets in Space Age around different stars can detect each other \\	
		\bottomrule
	\end{tabularx}	
	\begin{tablenotes}
		\footnotesize
		\item[a] Since K-type dwarfs are considered less extreme than M-type dwarfs, their extinction multiplier is set to be half that of M-type dwarfs.
	\end{tablenotes}	
\end{threeparttable}
\end{table*}

\section{Results and Discussion} \label{Section:Questions}
In this section we make numerical estimates of the model parameters and analyze the predictions made by the model. As we will see shortly, the Contact Probability can vary quite significantly depending on how the parameters are chosen. While we do not know the ``true" choice of our model parameters (the key obstacle all attempts to solve the Fermi Paradox have faced), we can address two important questions:
\begin{enumerate}
	\item What can we learn about the Contact Probability if we assume Earth to be a typical habitable planet in the galaxy?
	\item What can we learn about the Contact Probability if we let all model parameters vary randomly?
\end{enumerate}
First, however, we want to illustrate the model by running a few numerical examples. It is important to note that the parameters in our model are not point estimates, but rather represent averages of underlying distributions (in the case of the stage durations) or stage transition probabilities (in the case of the extinction rates). In fact, since stage durations follow an exponential distribution, where the standard deviation equals the mean, choosing any numerical value still allows for a very substantial flexibility. 

\subsection{Example Scenarios}\label{Section:Example}
For a few example scenarios, we will try to estimate the model parameters based on what we know about Earth's evolution.
\subsubsection{Stage Durations}
The stage durations describe how long evolution remains in a specific stage before transitioning to another. If we took Earth as a representative planet, with an age estimated at 4.5 billion years and the fossil record dating back nearly 4 billion years, the ``No Life" stage would last for a mere 500 million years. Based on the (geologically very recent) evolution of human intelligence, we want to estimate that Earth remained in the ``Primitive Life" stage for 4 billion years, and has been in the ``Intelligent Life" stage for only roughly 1 million years. Thus we set:
\begin{eqnarray*}
	&& w_0 = 5\cdot 10^8\ \text{years} \\
	&& w_1 = 4\cdot 10^9\ \text{years} \\
	&& w_2 = 1\cdot 10^6\ \text{years.}
\end{eqnarray*}
Since Space Age is ongoing on Earth, we don't have a measurement of its expected duration. But we have to acknowledge that even in the absence of any major global catastrophe, the continuation of Space Age on Earth has been all but certain, since space exploration efforts, such as the operation of radio telescopes, are subject to continued approval of private or public funding. In addition, it is only a fraction of Earth's total population that has developed the economic power to entertain space exploration. Even the most advanced civilizations in history have rarely lasted more than a millennium, and hence societies currently operating space telescopes may have a similar fate. When extrapolating beyond Earth, a point could also be made that, from a parsimony perspective, it would be reasonable to assume that the ``laws" of history apply across the entire galaxy, just as we naturally assume the laws of physics do. In other words, why should we assume that Space Age civilizations on other planets would perform better at maintaining their advanced state, rather than waging war among themselves over rather trivial disputes (which we have done, more or less, throughout the entirety of our existence)? Altogether, it seems fair to assume that Space Age should be a rather fragile stage, and for the purpose of this example scenario we want to set its average duration to  

\begin{equation*}
	w_3 = 1\cdot 10^3\ \text{years.}
\end{equation*}

\subsubsection{Extinction Rates} \label{Section:AbundantLife_R}
We have to keep in mind that the extinction rates do not represent the probability of whether one stage transitioned to another, but rather the probability of one stage transitioning to another, provided that it did transition at all. The probabilities of the ``whether" (or more precisely the ``when") are encoded in the stage durations, which we estimated above.\\

Unfortunately, the fact that life on Earth successfully advanced from the ``Primitive" to the ``Intelligent" stage, rather than going extinct, does not allow us to make any meaningful deductions on how probable this transition was. Since the onset of the Cambrian era several global catastrophes (super volcanism, meteorite strikes, major climate change etc.) have occurred and led to mass extinctions, but never eradicated life entirely. While we technically cannot rule out that primitive life on Earth did experience total extinction(s) and required several attempts to finally take hold, it seems fair to assume that life overall proved rather resilient on Earth. Thus, we assume a complete annihilation to be an improbable event and set:
\begin{eqnarray*}
	&& \alpha = 0.1 \\
	&& \beta = 0.1 \\
	&& \gamma = 0.1
\end{eqnarray*}

\subsubsection{Remaining Parameters}
As discussed above, it is widely believed that life around K- and M-type stars would face additional constraints, but it is open to speculation how tight such constraints would be. For the purpose of this  example we want to increase the rate of extinction around Red Dwarfs by a moderate amount and set:
\begin{equation*}
	M = 10
\end{equation*}
In terms of habitability, we acknowledge that alien lifeforms may thrive in conditions outside our current scientific understanding. But if we use Earth as a guidance, we can estimate this parameter based on the abundance of Earth-like exoplanets, for which there are relatively good estimates. While there is certainly variability in the literature \cite{fernandes_are_2025}, we want to apply a conservative estimate and assume that 20\% of stars across all types will harbor a habitable planet:
\begin{equation*}
	h = 20\%
\end{equation*}
Finally, we need to estimate a Detection Radius. For this we want to deviate from Earth as our guidance, since, if extraterrestrial signals were similar to Earth's in terms of type and signal strength (e.g.\ radio leakage), the resulting radius would in fact be very short, certainly under 10 lightyears. Contact Probabilities under these conditions would be insignificantly small ($<0.001$), which is an intuitive result given the very limited number of planets within this distance. In fact, the Fermi Paradox only poses a non-trivial problem if we assume a large enough Detection Radius. On the other hand, a vast Detection Radius would imply non-parsimonious assumptions as discussed earlier (Section \ref{Section:DetectionRadius}). Thus, as a certainly arbitrary but generous choice, we may set the Detection Radius to
\begin{equation*}
	D=500\ \text{lightyears.}
\end{equation*}

\subsubsection{Model Output and Discussion}
We can now analyze the model, and compute the probability for a planet to be in each of the four evolutionary stages for any given time. We see that with the parameters chosen above, life should be abundant. The model predicts that habitable planets around most star types should harbor life by the middle of their lifespan. Probability of life peaks around G-type stars (Table \ref{Tab:p_Midlife}), and declines slightly in Red Dwarfs (K- and M-type), likely as a consequence of the increased extinction rates. Planets around O- and B-type stars have the lowest probability of life, in accordance with the short lifespan of their hosting stars. 

\begin{table*}[htbp]
	\caption{Probability of planets around each star type to be in each of the stages in the model, ``No Life (stage 0)", ``Primitive Life (stage 1)", ``Intelligent Life (stage 2)", or ``Space Age (stage 3)", with model parameters chosen in section \ref{Section:Example}. Mathematically, $\text{P}[X_t=i], i=0,1,2,3$ with $t$ set to the middle of the host star's lifespan.}
	\label{Tab:p_Midlife}
	\begin{tabularx}{\linewidth}{L *{7}{C}}
		\toprule
		& O & B & A & F & G & K & M \\
		\midrule
		Mid-Life [years] & $2\cdot10^6$ & $8\cdot10^6$ & $4\cdot10^8$ & $2\cdot10^9$ & $6\cdot10^9$ & $1\cdot10^{10}$ & $4\cdot10^{11}$ \\
		No Life & 0.996 & 0.985 & 0.452 & 0.002 & 0.001 & 0.004 & 0.006 \\
		Primitive Life & 0.004 & 0.002 & 0.548 & 0.976 & 0.987 & 0.956 & 0.938 \\
		Intelligent Life & $<0.001$ & $<0.001$ & 0.001 & 0.001 & 0.001 & 0.001 & 0.001 \\		
		Space Age & $<10^{-5}$ & $<10^{-5}$ & $<10^{-5}$ & $<10^{-5}$ & $<10^{-5}$ & $<10^{-5}$ & $<10^{-5}$ \\
		\bottomrule
	\end{tabularx}
\end{table*}

\begin{figure}[htbp]
	\centering
	\includegraphics[width=1.0 \linewidth]{"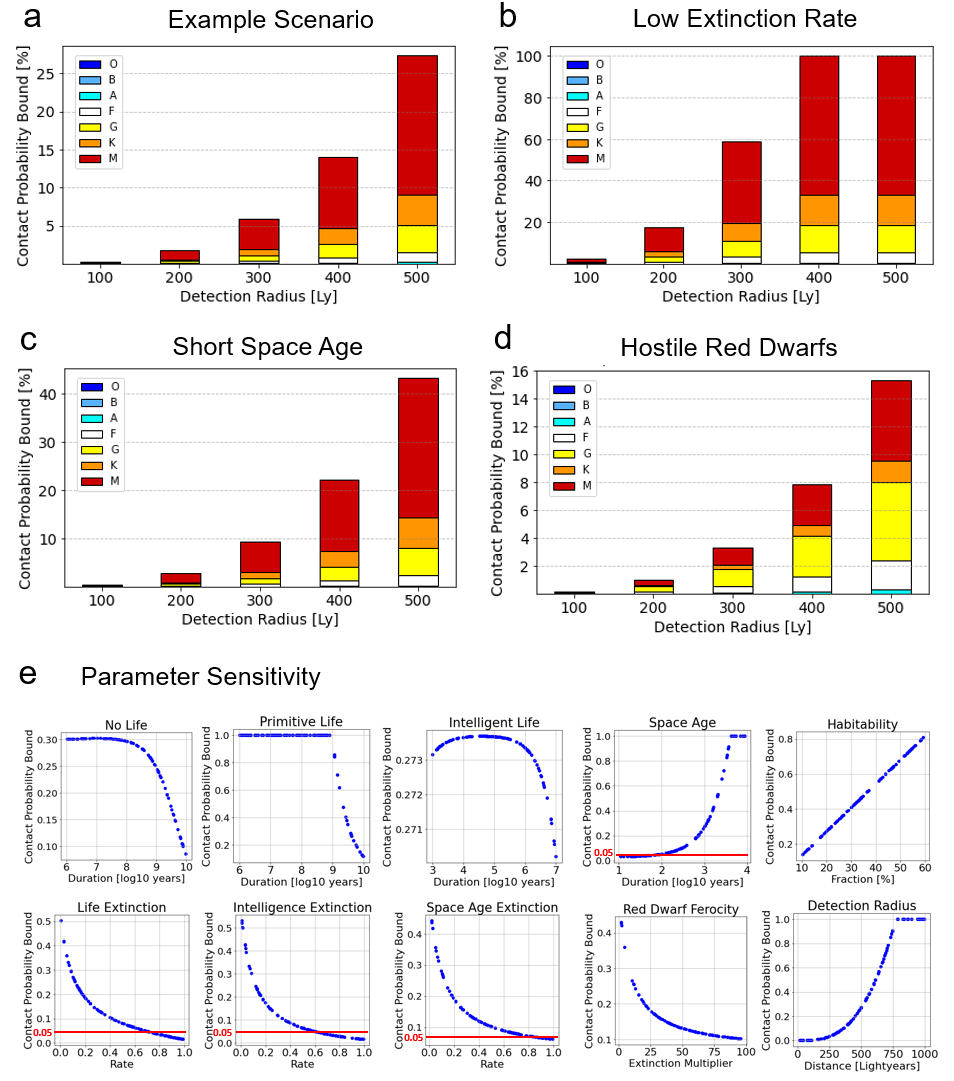"}
	\caption{Contact Probability under different parameter choices. \textbf{a:} Parameters chosen based on the example scenario in section \ref{Section:Example}. \textbf{b:} Reduced Extinction Rates ($\beta=\gamma=0.01$). \textbf{c:} Short Space Age duration ($\beta=\gamma=0.01$, $w_3=100$). \textbf{d:} Extremely hostile conditions around K- and M-type stars ($\beta=\gamma=0.01$, $w_3=100$, $M=100$). \textbf{e:} Contact Probability Bound with the indicated model parameters varying over a range. Only one parameter varies at a time. Remaining parameters left identical to the choices defined in section \ref{Section:Example}. The red line marks where Contact Probability falls under 0.05.}
	\label{Fig:Example}
\end{figure}

With these model parameters we obtain a Contact Probability of around 0.055 within a 300 lightyear Detection Radius and a Contact Probability of around 0.23 within a 500 lightyear Detection Radius (Table \ref{Tab:p_and_pfrak}). While these are arguably high enough numbers to consider the absence of any contact a paradox, we need to keep in mind that, with the exception of the stage durations $w_0, w_1, w_2$ and the habitability $h$, our parameter choices in section \ref{Section:Example} were mostly arbitrary. We therefore need to explore how our model predictions change as we try alternative choices. Before we proceed, we also note that the numerical computation of the Contact Probability $p$ via formula (\ref{Eqn:pC}) is rather cumbersome as the number of planets increase. Therefore, for the rest of the paper, we will use an approximation, as laid out in the Mathematical Appendix, Section \ref{Section:pBound}, to estimate an upper bound on the Contact Probability, the \textbf{Contact Probability Bound $\mathfrak{p}$}. The bound turns out to be fairly tight within the considered Detection Radii (Table \ref{Tab:p_and_pfrak}), and will allow for large-scale numerical exploration of the parameter space in Section \ref{Section:MonteCarlo}.

\begin{table*}[htbp]
	\caption{Contact Probabilities within the given Detection Radii and with the model parameters chosen in Section \ref{Section:Example}. For ease of computation, the Contact Probability Bound will serve as an upper bound estimate of the Contact Probability.}
	\label{Tab:p_and_pfrak}
	\begin{tabularx}{0.85\linewidth}{L *{5}{c}}
		\toprule
		Detection Radius [lightyears] & 100 & 200 & 300 & 400 & 500\\
		\midrule
		Contact Probability & 0.0021 & 0.0167 & 0.0548 & 0.1247 & 0.2288 \\
		Contact Probability Bound & 0.0022 & 0.0176 & 0.0592 & 0.1401 & 0.2734 \\
		\hline
	\end{tabularx}
\end{table*}

By considering the Contact Probability Bound, we can also easily see that the largest contribution of the ``probability mass" is coming from Red Dwarfs in our example scenario, as their high abundance sets off their reduced hospitality (Fig.~\ref{Fig:Example}a). The contribution of O- and B-type stars is negligible, in line with their low abundance as well as their short lifespan (Table \ref{Tab:StarTypes}). \\

To illustrate the sensitivity of the Contact Probability Bound to the model parameter choices, we could assume that intelligent life would overall be less likely to trigger a total apocalypse once it had reached intelligence, and set $\beta=0.01$ and $\gamma = 0.01$. In this scenario, we find that the Contact Probability Bound hits 1.0 (Fig.~\ref{Fig:Example}b), and thus contact could be close to certain. However, as we noted above, our assumption that the average civilization will be able to maintain their Space Age for 1,000 years may be too optimistic. Reducing the average Space Age duration to $w_3 = 100$ years reduces the Contact Probability Bound to 0.43 (Fig.~\ref{Fig:Example}c). We also may have been too optimistic about life around Red Dwarfs. Increasing the Red Dwarf Ferocity Factor to $M=100$, thereby making it significantly more likely for life to go extinct rather than evolve around these stars, we see that the Contact Probability Bound drops to just under 0.16. We also notice that, in this scenario, G-type stars now make the largest probability contribution (Fig.~\ref{Fig:Example}d). \\

In summary, we observe from these heuristic example runs that, even under rather pessimistic assumptions (inhospitable Red Dwarfs and short Space Age durations), the model predicts moderately high Contact Probability Bounds, reiterating the question of why such contact has not occurred.\\

However, while our parameter choices in this section may be reasonable, they are certainly still arbitrary, given the absence of any hard evidence for most of them. Every model parameter (with the exception of Intelligent Life Duration) significantly affects the Contact Probability Bound as can be seen from model runs with one parameter varying over a wide range of numerical choices (Fig.~\ref{Fig:Example}e). Thus, we cannot deduce meaningful answers to the question of probability of extraterrestrial contact based on any single numerical parameter choice and want to explore the parameter space by random sampling instead. 

\subsection{Monte Carlo Simulation}\label{Section:MonteCarlo}
\subsubsection{Parameter Range}
In this section we run the Markov model with the model parameters drawn from random distributions (Table \ref{Tab:MonteCarloParams}). While the ranges considered are certainly arbitrary, they still cover a broad spectrum of cosmological scenarios:\\

\begin{table*}[h]
	\caption{Model parameters and their numerical ranges considered.}
	\label{Tab:MonteCarloParams}	
	\begin{tabularx}{0.9\linewidth}{L *{3}{c}}
		\toprule
		Parameter & Monte Carlo Range & Example Scenario & Unit\\
		\midrule
		No Life Duration & $1\cdot 10^6$ to $1\cdot 10^{10}$ & $5\cdot 10^8$ & years \\
		Primitive Life Duration & $1\cdot 10^6$ to $1\cdot 10^{10}$ & $4\cdot 10^9$ & years \\
		Intelligent Life Duration & $1\cdot 10^3$ to $1\cdot 10^7$  & $1\cdot 10^6$ & years \\
		Space Age Duration & $1\cdot 10^1$ to $1\cdot 10^4$  & $1\cdot 10^3$ & years \\
		Life Extinction Rate & 0 to 1 & 0.1 & - \\
		Intelligence Extinction Rate & 0 to 1 & 0.1 & - \\
		Space Age Extinction Rate & 0 to 1 & 0.1 & - \\
		Red Dwarf Ferocity & 1 to 100 & 10 & - \\
		\bottomrule
	\end{tabularx}
\end{table*}

\textbf{No Life Duration ($w_0$)}: We let this parameter vary between 1 million and 10 billion years, following a log-uniform distribution, i.e. $w_0=10^x$ with $x$ being chosen from a uniform distribution in the closed interval [6.0, 10.0].\\
\textbf{Primitive Life Duration ($w_1$)}: We let this parameter vary between 1 million and 10 billion years, following a log-uniform distribution, i.e. $w_1=10^x$ with $x$ being chosen from a uniform distribution in the closed interval [6.0, 10.0]. \\
\textbf{Intelligent Life Duration ($w_2$)}: We let this parameter vary between 1,000 and 10 million years, following a log-uniform distribution, i.e. $w_2=10^x$ with $x$ being chosen from a uniform distribution in the closed interval [3.0, 7.0].\\
\textbf{Space Age Duration ($w_3$)}: We let this parameter vary between 10 and 10,000 years, i.e. $w_3=10^x$ with $x$ being chosen from a uniform distribution over the closed interval [1.0, 4.0].\\
\textbf{Life Extinction Rate ($\alpha$)}: We draw $\alpha$ from a uniform distribution over the open interval (0.0, 1.0). \\
\textbf{Intelligence Extinction Rate ($\beta$)}: We draw $\beta$ from a uniform distribution over the open interval (0.0, 1.0).\\
\textbf{Space Age Extinction Rate ($\gamma$)}: We draw $\gamma$ from a uniform distribution over the open interval (0.0, 1.0) \\
\textbf{Red Dwarf Ferocity ($M$)}: We draw $M$ from a uniform distribution over the closed interval [1.0, 100.0]. \\

While the estimates on \textbf{Habitability ($h$)} vary depending on dataset and methods \cite{fernandes_are_2025}, this parameter is more constrained than the others. We will therefore run the Monte Carlo simulation on the 8-dimensional space spanned by the parameters in Table \ref{Tab:MonteCarloParams}, with different fixed choices for $h$, varying between 10-60\%, covering the most accepted range of current estimates. \\
We will follow the reasoning discussed above (Section \ref{Section:Example}), and set the \textbf{Detection Radius ($D$)} to 500 lightyears, acknowledging that the Contact Probability Bounds estimated in the following sections would scale up if this parameter was increased.

\subsubsection{The Fermi Space}
Each parameter configuration in the Monte Carlo simulation represents one possible cosmological \textbf{scenario}, in which the durations of evolutionary eras and extinction rates are represented by the specific values of $w_0, w_1, w_2, w_3, \alpha, \beta, \gamma, M$, which the Markov model translates to a Contact Probability Bound $\mathfrak{p}$, i.e. the maximum probability of having been contacted within the ongoing Space Age on Earth. With this framework, the Fermi Paradox can now be phrased as: ``If we don't live in a scenario where the Contact Probability Bound is insignificantly small, why has contact not happened?" \\

While there is certainly ambiguity regarding what ``insignificantly small" means, we can adopt a commonly used threshold in statistical hypothesis testing and define Contact Probability as \textbf{"subcritical" if $\mathfrak{p}$ equals 0.05 or less}. Consequently, a parameter configuration, in which Contact Probability is subcritical, is a scenario, in which the absence of contact would be the expected outcome and, consequently, the Fermi Paradox would not hold. Thus, if we think of each parameter configuration as a point in an 8-dimensional space, then each point, for which $\mathfrak{p}\leq 0.05$ represents a solution to the Fermi Paradox. We want to call these points \textbf{Fermi Points}, and the set of all these points the \textbf{Fermi Space}.\\

In Fig.~\ref{Fig:Example}e, we can see that the Fermi Space contains solutions to the Fermi Paradox, in which, for example, Space Age on average only lasts for a few decades, or in which advanced life is overall too rare in the universe because of high extinction rates at any of its stages (approx. 0.7 or higher).
Given the diversity of all possible solutions to the Fermi Paradox, an interesting question is whether the Fermi Space has any discernible structure, i.e. whether its points fall into separate clusters (for example ``Life takes too long to evolve", or ``Space Age does not last long enough"). \\

\begin{figure}[htbp]
	\centering
	\includegraphics[width=1.0 \linewidth]{"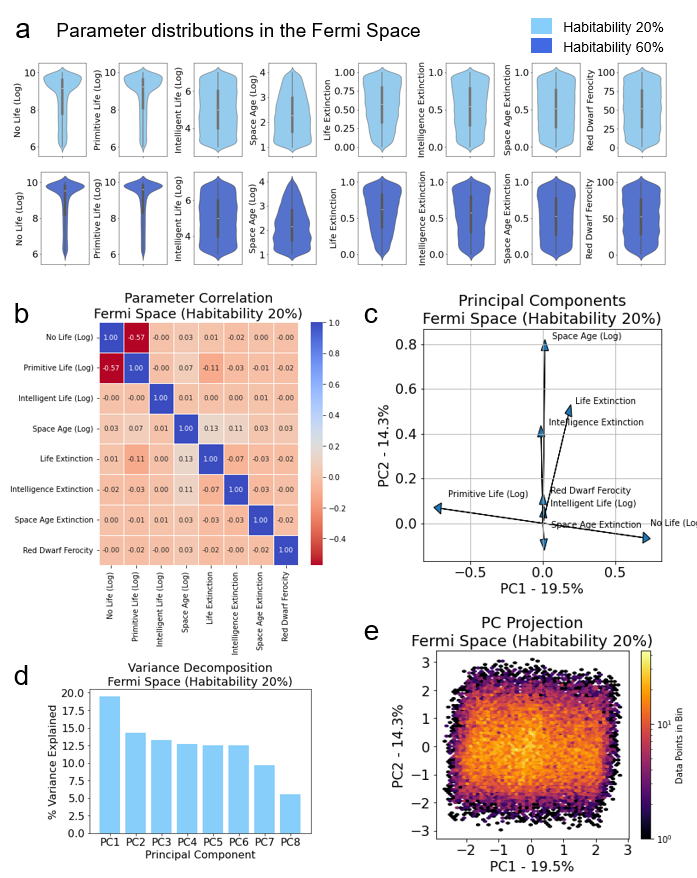"}
	\caption{\textbf{a:} Distribution of model parameters in the Fermi Space. Durations shown in logs of 10 years. \textbf{b:} Spearman correlation coefficients between parameters in the Fermi Space. \textbf{c:} Contributions of model parameters to the first two principal components in the Fermi Space. \textbf{d:} Percentage of variance explained by each principal component. \textbf{e:} Projection of the Fermi Space onto the first two principal components.}	
	\label{Fig:MC_PCA}
\end{figure}

Indeed, from a total of $10^5$ parameter configurations, we observe that within the Fermi Space the distributions of both No Life Duration and Primitive Life Duration are heavily skewed towards the high end of the range (Fig.~\ref{Fig:MC_PCA}a). This shows that solutions to the Fermi Paradox include a large number of scenarios, in which life on average takes a long time---more than 1 billion years---to emerge and to advance to intelligence. Life Extinction Rate also shows a skew towards the high end of its range, showing that solutions to the Fermi Paradox tend to be scenarios, in which catastrophic extinctions of all life on a planet are more frequent events. Conversely, Space Age Duration is skewed towards the low end of the range, indicating that solutions to the Fermi Paradox tend to be scenarios, in which advanced civilizations are rather short-lived. The remaining parameters show only very minor skew, indicating that solutions to the Fermi Paradox can sit anywhere within their respective range. With increasing Habitability, the skew across all variables increases (Fig.~\ref{Fig:MC_PCA}a). This is intuitive since the Contact Probability Bound scales with Habitability (Fig.~\ref{Fig:Example}e), and thus more adverse conditions in the remaining model parameters are required to drive Contact Probability below the subcritical threshold if Habitability is high.\\

We also see that within the Fermi Space, No Life Age Duration and Primitive Life Age Duration are anti-correlated, indicating that in order to produce a subcritical Contact Probability, a short duration of the one requires a long duration of the other (Fig.~\ref{Fig:MC_PCA}b). Neither of these stage durations correlate with Intelligent Life Duration, which appears counter-intuitive, but is likely explained by the shorter duration of the latter in our model, reducing its ability to counteract the other two stages, which may outlast Intelligent Life by several orders of magnitude. Likewise, we observe that the first principal component of the Fermi Space shows a high positive contribution of No Life Duration and a high negative contribution of Primitive Life Age Duration, indicating that the highest variance in the Fermi Space happens along an axis where short duration of the one is offset by a high duration of the other (Fig.~\ref{Fig:MC_PCA}c). Otherwise, principal component analysis reveals little structure in the Fermi Space, with the variance being split fairly evenly, and the projection being approximately concentric (Fig.~\ref{Fig:MC_PCA}d,e). We therefore conclude that the solutions to the Fermi Paradox seem to form a continuum in the parameter space and do not fall into pronounced clusters.\\ 

\subsubsection{Dimensionality Reduction}\label{Section:Simplification}
For a better visualization of the Fermi Space and its structure, we want to reduce dimensionality manually and combine No Life Duration, Primitive Life Duration, and Intelligent Life Duration into a single parameter, representing the total time evolution takes on a planet until Space Age is reached. In addition, we combine the stage-specific extinction rates $\alpha$, $\beta$, and $\gamma$ into a single parameter, summarizing the total risk of extinction, irrespective of stage. \\

Thus, the transformed Monte Carlo dataset is now only 3-dimensional and contains the following parameters:
\begin{enumerate}
	\item The combined durations of the No Life, Primitive Life, and Intelligent Life stages $W:=w_0+w_1+w_2$. We call this parameter \textbf{Emergence Time}.
	\item \textbf{Space Age Duration $w_3$}.
	\item The total of Life Extinction Rate\footnote{Since the Life Extinction Rate $\alpha$ depends on star type, we build an aggregated parameter: $\alpha_x = \zeta_M\cdot\text{min}(1,M\alpha) + \zeta_K\cdot\text{min}(1,K\alpha) + (1-\zeta_M-\zeta_K)\cdot\alpha$, with $\zeta_M, \zeta_K$ being the abundance of M- and K-stars.}, Intelligence Extinction Rate, and Space Age Extinction Rate $E := \alpha_x+\beta+\gamma$. We call this parameter the \textbf{Extinction Rate}.
\end{enumerate}

\begin{figure}[htbp]
	\centering
	\includegraphics[width=1.0 \linewidth]{"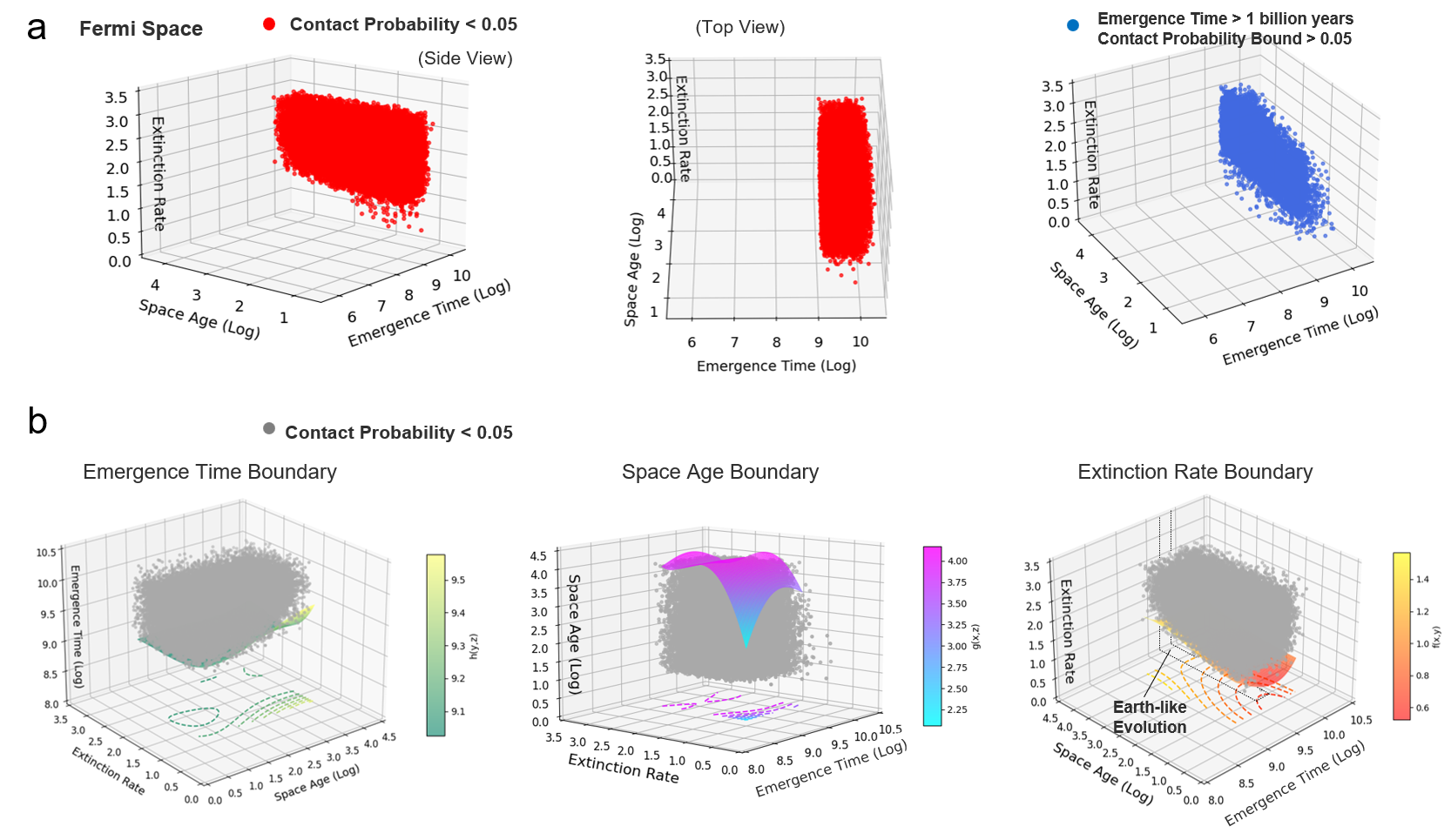"}
	\caption{\textbf{a:} \textit{left/center:} 3D projection of the Fermi Space, i.e. all points in the Monte Carlo dataset that result in a Contact Probability Bound of less than 0.05 (Fermi Points). \textit{right:} 3D projection of points in the Monte Carlo simulation that result in a Contact Probability Bound greater than 0.05 (Non-Fermi Points). Only Non-Fermi Points with Emergence Time longer than 1 billion years are shown. \textbf{b:} 3D projection of the Fermi Space with estimated envelope planes marking the boundary, beyond which points lie outside the Fermi Space. The envelope planes are generated by binning the x-/y-plane into 10x10 segments and connecting the minimal (Extinction Rate, Emergence Time) or maximal (Space Age) points in these bins using forth-order polynomial regression curves. ``Earth-like Evolution" (right) refers to scenarios, in which Space Age Duration exceeds 100 years and Emergence Time is approximately 4.5 billion years (bin: 4.1-6.5 billion years). Habitability set to 20\%.} 	
	\label{Fig:MC_FermiSpace}
\end{figure}

With this dimensionality reduction, the Fermi Space can now be visualized as a point cloud (Fig.~\ref{Fig:MC_FermiSpace}a), and we can see that it appears to have rather sharp boundaries, especially along the Emergence Time axis, which we can numerically estimate using envelope functions (Fig.~\ref{Fig:MC_FermiSpace}b). Based on the contour lines of these envelope functions, we can now observe the following interesting phase transitions in the parameter space: 
\begin{itemize}  
	\item There are no Fermi Points lower than $10^9$ on the Emergence Time axis. Thus, scenarios, in which life evolves to Space Age in less than roughly 1 billion years on average, would automatically pose a Fermi Paradox (Fig.~\ref{Fig:MC_FermiSpace}a, top view).
	\item There are Fermi Points across the entire Space Age axis. Thus, even scenarios, in which civilizations maintain space exploration for as long as 10,000 years on average, can still lead to subcritical Contact Probabilities, i.e. a long Space Age does not automatically pose a Fermi Paradox. 
	\item There are no Fermi Points lower than roughly 0.6 on the Extinction Rate axis. Thus, scenarios, in which life does not have at least a moderate chance of extinction at any of the transitions in its evolution, automatically pose a Fermi Paradox. 
\end{itemize}

It is important to note that these boundaries are visualizations of a projected dataset. Thus, the envelope functions only represent one-way boundaries, i.e. boundaries outside of which there are no Fermi Points. In fact, non-Fermi Points can lie adjacent to Fermi Points inside of the boundaries although they are more sparse (Fig.~\ref{Fig:MC_FermiSpace}a, right), and hence the inversions of the above observations are not necessarily true. For example, we cannot conclude that all scenarios, in which Emergence Time lasts longer than 1 billion years, automatically do solve the Fermi Paradox. The question of how ``likely" the Fermi Paradox holds is equivalent to the question of how abundant Fermi Points are among all considered scenarios, either across the entire dataset or in subsets of interest, for example in the subset of scenarios resembling a ``Earth-like" Evolution (Fig.~\ref{Fig:MC_FermiSpace}b, right). We will address this question in the next section.

\subsubsection{Contact Probability Distribution} 
\begin{figure}[htbp]
	\centering
	\includegraphics[width=1.0 \linewidth]{"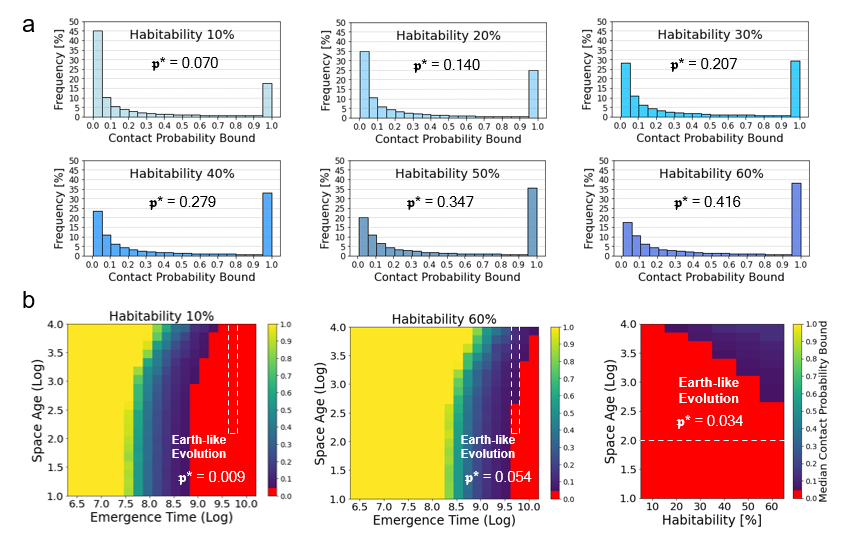"}
	\caption{\textbf{a:} Distribution of Contact Probability Bounds from $10^5$ simulations with random parameter configurations and fixed Habitability. \textbf{b:} \textit{left/center:} Color-coded Median Contact Probability Bound over bins of Emergence Time and Space Age Duration, with Habitability set to either 10\% or 60\%. The red area marks Emergence Time / Space Age combinations where the majority of all possible Extinction Rates result in a subcritical Contact Probability. The dashed lines marks scenarios resembling an Earth-like evolution (Emergence Time bin: 4.1-6.5 billion years, Space Age $>$ 100 years).  \textit{right:} Earth-like evolution bins taken from heatmaps with different choices for Habitability. $\mathfrak{p^*}$: median Contact Probability Bound.}	
	\label{Fig:MC_pHisto}
\end{figure}

Across the Monte Carlo dataset, we see that the resulting distribution of Contact Probability Bounds follows a bimodal shape, with the mass largely concentrated in both ends of the range. For example, if we assume a Habitability of 10\%, Fermi Points make up 45\% of the entire dataset (Fig.~\ref{Fig:MC_pHisto}a), putting the median Contact Probability Bound to under 0.07 (Table \ref{Tab:Median_p}). Even when assuming a Habitability of 60\%, we obtain a median Contact Probability Bound of only 0.416, even if we do not constrain any other model parameters (Fig.~\ref{Fig:MC_pHisto}a, Table \ref{Tab:Median_p}).\\

Since we considered a wide numerical range for the parameters in our Monte Carlo simulation, the absence of any constraints will certainly retain a large number of rather exotic scenarios, for example ones in which life evolves several orders of magnitude faster than it did on Earth. Thus, if we now want to return to our original question, and wonder what predictions our model can make if we assumed that Earth was a typical habitable planet in the local galaxy, we have to constrain the parameter space such that we retain only scenarios where the evolutionary trajectory resembles an "Earth-like Evolution", which we define as follows (also see Fig.~\ref{Fig:MC_FermiSpace}b right, black dotted lines):

\begin{enumerate}
	\item Emergence Time is approximately 4.5 billion years,
	\item Space Age can last between 100 and 10,000 years,
	\item The Extinction Rate is unconstrained.
\end{enumerate} 

Again we want to emphasize that these numbers represent the means of random distributions rather than point estimates, meaning that ``fixing" these parameters at numerical values still allows for a large diversity of scenarios.\\

For a more intuitive graphical visualization, we can segment the Emergence Time / Space Age Duration plane into bins and average the Contact Probability Bounds over the Extinction Rate axis, essentially creating a 2D representation of the parameter space. We notice that the median Contact Probability Bound gradually decreases as Emergence Time increases until it reaches the subcritical threshold, the exact crossing of which depends on Space Age Duration: The longer the Emergence Time, the longer need civilizations remain in Space Age to avoid subcritical Contact Probability (Fig.~\ref{Fig:MC_pHisto}b). We can easily observe that with a Habitability of 10\%, scenarios resembling Earth-like evolution produce predominantly subcritical Contact Probability. Increasing Habitability to 60\% shifts the distribution on the Emergence Time axis, shrinking the subcritical area, but the bracket representing Earth-like evolution still contains a substantial amount of scenarios with subcritical Contact Probability, putting the median to 0.054 (Fig.~\ref{Fig:MC_pHisto}b, center). Aggregating over all choices of Habitability considered in our dataset, we obtain a median Contact Probability Bound of only 0.034 (Fig.~\ref{Fig:MC_pHisto}b, right; Table \ref{Tab:Median_p}). Consequently, if Earth was a typical planet, we should expect an overall low probability of having received any extraterrestrial signals throughout the past decades.

\begin{table*}[h]
	\caption{Contact Probability Bound under different scenarios. Median from $1\cdot 10^5$ (1,2,4,5) or $6\cdot 10^5$ (3,6) random parameter configurations.}
	\label{Tab:Median_p}
	\begin{tabularx}{0.8\linewidth}{c L c c}
		\toprule
		& Scenario set & Habitability & Contact Probability \\
		\midrule
		1 & (No Assumptions) & 10\% & $<0.070$ \\
		2 & (No Assumptions) & 60\% & $<0.416$ \\
		3 & (No Assumptions) & 10-60\% & $<0.213$ \\
		4 & Earth-like Evolution & 10\% & $<0.009$ \\
		5 & Earth-like Evolution & 60\% & $<0.054$ \\
		6 & Earth-like Evolution & 10-60\% & $<0.034$ \\
		\bottomrule
	\end{tabularx}
\end{table*}

\section{Conclusion}
In summary, based on simulation of a large number of parameter configurations, our Markov Model predicts overall strikingly low probabilities of contact with other civilizations, in particular if we assume that conditions on Earth are at least somewhat representative across habitable planets in the local galaxy. Ongoing observations from the JWST and future missions will continue to gather valuable data on atmospheres and conditions on exoplanets in our neighborhood, which will help constrain the Habitability parameter in our model and sharpen its predictive power. Other possible refinements to the presented model could involve the implementation of Poisson processes to describe the spatiotemporal distribution of exoplanets, rather than relying on a uniform approximation, in order to account for variations in both local density and star formation events. \\
It is interesting to note that various other mathematical approaches have independently arrived at similar results to those of our model, concluding that, while primitive life would likely be common \cite{veres_solitude_2026}, advanced life likely is not \cite{kipping_objective_2020,sandberg_dissolving_2018}, or we are at least unlikely to detect it \cite{grimaldi_signal_2017,grimaldi_inferring_2023}, calling into question whether the Fermi Paradox truly qualifies as a paradox. These conclusions should not be interpreted as an argument against the SETI program or the search for extraterrestrial intelligence more broadly. Rather, they suggest that the paradox arises only under a specific set of assumptions, which seem to be a lot less robust than is often appreciated. We can forgive Dr. Fermi---brilliant as he was---for not running the math in real time while pondering his famous question over a sandwich. But mathematically, contact with E.T. does not seem to have been so likely after all.

\section{Author Contributions}
\textbf{Philipp N. Spahn} conceptualized the project, designed the model, drafted the mathematics, wrote the Python implementation, ran analyses, and wrote the manuscript. \textbf{Elmar Teufl} led the mathematics, edited the manuscript, and provided critical feedback and advice.

\section{Acknowledgments}
We would like to thank Guangwei Fu for critically reading the draft, providing helpful advice, and for endorsing the manuscript for publication on arXiv. We would also like to thank Isaac Shamie for helpful comments and suggestions.

\section{Mathematical Appendix}\label{Section:mathappendix}

\subsection{Basic Markov Chain Theory}\label{Section:basicmarkov}
In the following, we quickly review basic results from Markov Chain theory: Let 
\begin{equation*}
	W_t := \text{inf}\{\sigma>0:\ X_{t+\sigma}\neq X_t\}	
\end{equation*}
be the stage durations for $X$, i.e. the time the process spends in one stage before jumping to another. Stage durations follow an exponential distribution with ``escape rates" $q_i$, i.e. the average time the process spends in stage $i$ is given by:
\begin{equation}\label{Eqn:waittime}
	w_i:=\text{E}[W_t|X_t=i] = \frac{1}{q_i}
\end{equation}
The process is completely characterized by its generator matrix $Q = (q_{ij})_{0\leq i,j \leq 3}$, which comprises the information encoded in both the jump matrix and the escape rates:  
\begin{eqnarray}\label{Eqn:qmatrix_computation}
	&& q_{ij} = q_i\cdot r_{ij}\ \text{for}\ i\neq j\\
	&& q_{ii} = -q_i \nonumber
\end{eqnarray}
We assume that $Q$ is diagonalizable and let $\lambda = (\lambda_0,\lambda_1,\lambda_2,\lambda_3)$ be its eigenvalues. Let $A$ be a transformation matrix such that
\begin{equation*}
	A^{-1} Q A = 
	\begin{bmatrix}
		\lambda_0 & 0 & 0 & 0 \\
		0 & \lambda_1 & 0 & 0 \\
		0 & 0 & \lambda_2 & 0 \\
		0 & 0 & 0 & \lambda_3
	\end{bmatrix}
\end{equation*}
The transition matrices
\begin{equation}\label{PiDef}
	\Pi(t)=(\pi_{ij}(t))_{0\leq i,j \leq 3}:=(\text{P}[X_t=j|X_0=i])_{0\leq i,j \leq 3}
\end{equation}
describe the probabilities of jumping from stage $i$ to stage $j$ in time $t$. We note that $X$ is time-stationary, i.e. for any time shift $\tau_0\geq 0$:
\begin{equation}\label{stationary}
	\text{P}[X_{t+\tau_0}=j|X_{\tau_0}=i] = \text{P}[X_t=j|X_0=i] = \pi_{ij}(t)
\end{equation}
$\Pi(t)$ can be computed from the generator matrix $Q$ via:
\begin{equation*}
	\Pi(t) = e^{tQ} = A\cdot 
	\begin{bmatrix}
		e^{\lambda_0t} & 0 & 0 & 0 \\
		0 & e^{\lambda_1t} & 0 & 0 \\
		0 & 0 & e^{\lambda_2t} & 0 \\
		0 & 0 & 0 & e^{\lambda_3t}
	\end{bmatrix}
	\cdot A^{-1}
\end{equation*}
Thus, with $A = (a_{ij})_{0\leq i,j \leq 3}$ and $A^{-1} = (\hat a_{ij})_{0\leq i,j \leq 3}$ we can compute the transition probabilities for $t>0$:
\begin{eqnarray} \label{Eqn:ComputePi}	
	\lefteqn{\Pi(t) = } \nonumber\\ 
	&& = \begin{bmatrix}
		e^{\lambda_0t}
		\begin{pmatrix}
			a_{00} \\
			a_{10} \\
			a_{20} \\
			a_{30}
		\end{pmatrix}
		e^{\lambda_1t}
		\begin{pmatrix}
			a_{01} \\
			a_{11} \\
			a_{21} \\
			a_{31}
		\end{pmatrix}
		e^{\lambda_2t}
		\begin{pmatrix}
			a_{02} \\
			a_{12} \\
			a_{22} \\
			a_{32}
		\end{pmatrix}
		e^{\lambda_3t}
		\begin{pmatrix}
			a_{03} \\
			a_{13} \\
			a_{23} \\
			a_{33}
		\end{pmatrix}
	\end{bmatrix}
	\cdot A^{-1} \nonumber\\
	&&= \begin{bmatrix}
		\sum_r e^{\lambda_rt}a_{0r}\hat a_{r0} & \dots & \sum_r e^{\lambda_rt}a_{0r}\hat a_{r3} \\
		\vdots & \dots & \vdots \\
		\sum_r e^{\lambda_rt}a_{3r}\hat a_{r0} & \dots & \sum_r e^{\lambda_rt}a_{3r}\hat a_{r3}
	\end{bmatrix} \nonumber\\
	&&=
	\left(\sum_{r=0}^3 e^{\lambda_rt}a_{ir}\hat a_{rj}\right)_{0\leq i,j \leq 3}
\end{eqnarray}
Since all row sums of $Q$ are zero, we know the first eigenvalue/eigenvector pair is $\lambda_0 = 0$ and $(a_{00},a_{10},a_{20},a_{30})=(1,1,1,1)$. But we note that the remaining eigenvalues may be complex. \\

We also note that, with stage durations spanning millions of years or more (as we will explore in the next session), $X$ reaches its stationary distribution typically only over very large time scales, and we therefore use direct computation of the transition probabilities throughout the paper.

\subsection{Contact Probability for a Single Exoplanet}\label{Section:P(Cn)}
We now want to compute $\text{P}(C_n)$, the probability to be contacted by a given exoplanet in distance $d$, which formed at a time $\tau_0$. As we stated in Section \ref{Section:ContactProb_Interval}, this probability equals the probability that this planet is in Space Age for some time during its Contact Interval:
\begin{equation*}
	\text{P}(C_n) = p_d(\tau_0) := \text{P}_{\tau_0}[X_\tau=3\ \textit{for some}\ \tau\in I_d]
\end{equation*}
with $\text{P}_{\tau_0}(\ \cdot\ ) := \text{P}(\ \cdot\ |\ X_{\tau_0}=0)$. Using the total probability theorem, we expand this to:
\begin{eqnarray*}
	&& = \sum_{k=0}^3 \text{P}_{\tau_0}[X_\tau=3\ \textit{for some}\ \tau\in I_d\ |\ X_{\tau_d}=k]\cdot \text{P}_{\tau_0}[X_{\tau_d}=k] \\
	&& = \sum_{k=0}^2 \underbrace{\text{P}_{\tau_0}[X_\tau=3\ \textit{for some}\ \tau\in I_d\ |\ X_{\tau_d}=k]}_{=:\ (*)} \cdot \underbrace{\text{P}[X_{\tau_d}=k\ |\ X_{\tau_0}=0]}_{\stackrel{(\ref{stationary})}=\ \pi_{0k}(\tau_d-\tau_0)} \\
	&& \ +\ \underbrace{\text{P}_{\tau_0}[X_\tau=3\ \textit{for some}\ \tau\in I_d\ |\ X_{\tau_d}=3]}_{=\ 1} \cdot\ \underbrace{\text{P}[X_{\tau_d}=3]\ |\ X_{\tau_0}=0]}_{\stackrel{(\ref{stationary})}=\ \pi_{03}(\tau_d-\tau_0)}
\end{eqnarray*}
with $\tau_d$ being the beginning of its Contact Interval (Fig.~\ref{Fig:StarFormContactInt}c). We can simplify the expression $(*)$ using basic Markov Chain properties, i.e. the independence of the past ($t=\tau_0$) as well as the time-stationarity (shift by $\tau_d$): 
\begin{equation*}
	(*) = \text{P}[X_\tau=3\ \textit{for some}\ \tau\in [0,S]\ |\ X_0=k]
\end{equation*}

To compute this expression, we construct an auxiliary process $\widetilde{X}^{(k)}$ that starts at $\widetilde{X}^{(k)}_0 = k$ and is absorbed in stage 3. Thus, its jump matrix is
\begin{equation*}
	\widetilde{R} = 
	\begin{bmatrix}
		0 & 1 & 0 & 0 \\
		\alpha & 0 & 1-\alpha & 0 \\
		0 & \beta & 0 & 1-\beta \\
		0 & 0 & 0 & 1
	\end{bmatrix}
\end{equation*}
$\widetilde{X}^{(k)}$ has the same escape rates as $X$, with the exception that $q_3 = 0$, since stage 3 is absorbing. Thus, the generator matrix is
\begin{equation*}
	\widetilde{Q} = 
	\begin{bmatrix}
		-q_0 & q_0 & 0 & 0 \\
		\alpha q_1 & -q_1 & (1-\alpha)q_1 & 0 \\
		0 & \beta q_2 & -q_2 & (1-\beta) q_2 \\
		0 & 0 & 0 & 0
	\end{bmatrix}
\end{equation*}
and we consider the submatrix:
\begin{equation*}
	\widehat{Q} := 
	\begin{bmatrix}
		-q_0 & q_0 & 0 \\
		\alpha q_1 & -q_1 & (1-\alpha)q_1 \\
		0 & \beta q_2 & -q_2  \\
	\end{bmatrix}
\end{equation*}
The expression $(*)$ then equals the probability that the process $\widetilde{X}^{(k)}$ takes less than $S$ years to become absorbed. Mathematically, we can express this using the absorption time $U^{(k)}:=\text{inf}\{t>0:\ \widetilde{X}^{(k)}_t=3\}$ as:
\begin{equation*}
	(*) = \Phi_{k,S}:= \text{P}[U^{(k)}\leq S]
\end{equation*} 
which can be shown to equal
\begin{equation*}
	\Phi_{k,S} = 1 - \langle \delta_k\ ,\ e^{S\widehat{Q}}\cdot\mathbf{1} \rangle
\end{equation*}
with $\delta_k$ being the deterministic start distribution in stage $k$ and $\mathbf{1} := (1,1,1)^T$ being the 1-vector.

Thus, we obtain the contact probability:
\begin{equation}\label{Eqn:PCn}
	\text{P}(C_n) = p_d(\tau_0) = \pi_{03}(\tau_d-\tau_0) + \sum_{k=0}^2 \Phi_{k,S} \cdot \pi_{0k}(\tau_d-\tau_0)
\end{equation}

\subsection{Contact Probability within a Given Detection Radius}\label{Section:p}
With Formula (\ref{Eqn:PCn}) for a single exoplanet, we can now compute the Contact Probability within the Detection Radius via Formula (\ref{Eqn:pC}):\\
\begin{equation*}
	p = 1-\prod_{d=1}^D\prod_{n=1}^{N_d} [1-\text{P}(C_n)]
\end{equation*}
We first fix a single distance of $d$ lightyears. As discussed in section \ref{Section:StarAgeAndPos}, there are $N_d$ habitable planets at this distance, with $N_d$ being a sum of habitable planets around stars of each type
\begin{equation*}
	N_d = \nu_d^O + \nu_d^B + ... + \nu_d^M
\end{equation*}
We set:
\begin{equation*}
	P_d := \prod_{n=1}^{N_d} [1-\text{P}(C_n)] =  \underbrace{\prod_{n=1}^{\nu_d^O} [1-\text{P}(C_n)]}_{=:P_d^O} \cdot ... \cdot \underbrace{\prod_{n=1}^{\nu_d^M}[1-\text{P}(C_n)]}_{=:P_d^M}
\end{equation*}

We first consider planets around stars of type O. As discussed in section \ref{Section:StarAgeAndPos}, we assume that these formed in discrete time intervals. Thus, going back in time from the beginning of their Contact Interval, we assume that each planet formed at:
\begin{equation*}
	\tau_0^{(n)} := \tau_d - L^O + n\cdot \Delta T_d^O
\end{equation*} 
with a $0\leq n \leq \nu_d^O-1$ (Fig.~\ref{Fig:StarFormContactInt}b). Thus we can compute:
\begin{eqnarray*}
	&& P_d^O=\prod_{n=1}^{\nu_d^O} [1-\text{P}(C_n)] = \prod_{n=0}^{\nu_d^O-1} [1-p_d(\tau_0^{(n)})] = \prod_{n=0}^{\nu_d^O-1} [1-p_d(\tau_d-L^O+n\cdot\Delta T_d^O)]\\
	&& \stackrel{(\ref{Eqn:DeltaT})} = \prod_{n=0}^{\nu_d^O-1} [1-p_d[\tau_d-(\nu_d^O-n)\cdot\Delta T_d^O]] = \prod_{n=1}^{\nu_d^O} [1-p_d(\tau_d-n\cdot\Delta T_d^O)]\\
	&& \stackrel{(\ref{Eqn:PCn})}= \prod_{n=1}^{\nu_d^O} [1 - \pi_{03}(n\cdot\Delta T_d^O) - \sum_{k=0}^2 \Phi_{k,S}\cdot \pi_{0k}(n\cdot\Delta T_d^O)]
\end{eqnarray*}
Thus, after analogous computation for the other star types, we have the Contact Probability:
\begin{equation*}
	p = 1-\prod_{d=1}^D P_d = 1-\prod_{d=1}^D P_d^O\cdot ... \cdot P_d^M
\end{equation*}

\subsection{Contact Probability Bound for a Single Exoplanet}
Direct computation of $\text{P}(C_n)$ via Formula (\ref{Eqn:PCn}) is numerically costly. We therefore want to make the following simplification: Since for relevant parameter choices the process $X$ moves on geological timescales, the probability to reach Space Age within just a few decades are typically negligibly small. We can therefore replace the absorption probabilities $\Phi_{k,S}$ in Formula (\ref{Eqn:PCn}) with a more tractable expression.

As in Section \ref{Section:P(Cn)} we have:
\begin{equation*}
	\text{P}(C_n) = \pi_{03}(\tau_d-\tau_0) + \sum_{k=0}^2 \underbrace{\text{P}[X_\tau=3\ \textit{for some}\ \tau\in [0,S]\ |\ X_0=k]}_{=:\ (*)}\cdot \pi_{0k}(\tau_d-\tau_0)
\end{equation*}
If the process hit stage 3 (starting in $k\neq 3$) during the interval $[0,S]$, then its stage duration in stage $k$ was obviously shorter than $S$ years, i.e.
\begin{equation*}
	[X_\tau=3\ \textit{for some}\ \tau\in [0,S],\ X_0=k]\subseteq [W_0\leq S,\ X_0=k]
\end{equation*}
and thus we obtain the estimate
\begin{eqnarray}\label{Eqn:PCn_Bound}
	\text{P}(C_n) && \leq \pi_{03}(\tau_d-\tau_0) + \sum_{k=0}^2\text{P}[W_0 \leq S\ |\ X_0=k]\cdot \pi_{0k}(\tau_d-\tau_0) \nonumber \\
	&& = \pi_{03}(\tau_d-\tau_0) + \sum_{k=0}^2 (1-e^{-q_k S})\cdot \pi_{0k}(\tau_d-\tau_0)
\end{eqnarray}

\subsection{Contact Probability Bound within a Given Detection Radius}\label{Section:pBound}
We are further going to use the following approximation for a more efficient numerical estimate of the total Contact Probability:
\begin{equation*}
	p:=\text{P}(C)=\text{P}(\bigcup_{n=1}^{N} C_n)\leq \sum_{n=1}^N \text{P}(C_n)
\end{equation*}
While this bound may appear rather loose at first glance, we note that for typical parameter choices the $C_n$ are rare events, so that their overlaps are typically negligible, making the above sum a reasonable approximation. Finally, since we can group the $N$ planets by distance, each group containing $N_d$ habitable planets, we have
\begin{equation}
	p \leq \sum_{d=1}^D \sum_{n=1}^{N_d} \text{P}(C_n)
\end{equation}

With the above estimate, Formula (\ref{Eqn:PCn_Bound}) for a single exoplanet, we can now derive an explicit formula for an upper bound on $p$.\\

Analogously to Section \ref{Section:p}, we first fix a single distance of $d$ lightyears, and group the $N_d$ habitable planets at this distance by star type:
\begin{equation*}
	N_d = \nu_d^O + \nu_d^B + ... + \nu_d^M
\end{equation*}
and consider the birth times $\tau_0$ of each planet around stars of each type (starting with O-type), assuming equidistant birth intervals:
\begin{equation*}
	\tau_0^{(n)} := \tau_d - L^O + n\cdot \Delta T_d^O
\end{equation*} 
with a $0\leq n \leq \nu_d^O-1$ (Fig.~\ref{Fig:StarFormContactInt}b). To simplify notation in the following we omit the 'O' index, but note that the variables in the calculation below depend on star type. By summing over all planets around O-type stars in distance $d$ lightyears we obtain:
\begin{eqnarray*}
	&& \sum_{n=1}^{\nu_d} \text{P}(C_n) = \sum_{n=0}^{\nu_d-1} p_d(\tau_0^{(n)}) = \sum_{n=0}^{\nu_d-1} p_d(\tau_d-L+n\cdot\Delta T_d)\\
	&& \stackrel{(\ref{Eqn:PCn_Bound})}\leq \sum_{n=0}^{\nu_d-1} \pi_{03}(L-n\cdot\Delta T_d) + \sum_{k=0}^2 \left(1-e^{-q_k S}\right)\cdot \pi_{0k}(L-n\cdot\Delta T_d)\\
	&& \stackrel{(\ref{Eqn:DeltaT})}= \sum_{n=0}^{\nu_d-1} \pi_{03}[(\nu_d-n)\cdot\Delta T_d] + \sum_{k=0}^2 \left(1-e^{-q_k S}\right)\cdot \pi_{0k}[(\nu_d-n)\cdot\Delta T_d]\\
	&& = \sum_{n=1}^{\nu_d} \pi_{03}(n\cdot\Delta T_d) + \sum_{k=0}^2 \left(1-e^{-q_k S}\right)\cdot \pi_{0k}(n\cdot\Delta T_d)\\
	&& \stackrel{(\ref{Eqn:ComputePi})}= \underbrace{\sum_{n=1}^{\nu_d} \sum_{r=0}^3 \left(e^{\lambda_r \Delta T_d}\right)^n a_{0r}\hat a_{r3}}_{=: (*)} + \underbrace{\sum_{n=1}^{\nu_d}\sum_{k=0}^2 \left(1-e^{-q_k S}\right)\sum_{r=0}^3 \left(e^{\lambda_r \Delta T_d}\right)^n a_{0r}\hat a_{rk}}_{=:(**)}	
\end{eqnarray*}
Changing the order of summation
\begin{equation*}
	(**)  = \sum_{k=0}^2 \left(1-e^{-q_k S}\right)\sum_{r=0}^3 a_{0r}\hat a_{rk}\sum_{n=1}^{\nu_d}\left(e^{\lambda_r \Delta T_d}\right)^n
\end{equation*}
and using the geometric sum and taking into account that $\lambda_0=0$ and $a_{00}=1$, this expression simplifies to:
\begin{eqnarray*}
	&& = \sum_{k=0}^2 \left(1-e^{-q_k S}\right) \biggl[ \nu_d\cdot \hat a_{0k} + \underbrace{\sum_{r=1}^3 a_{0r}\hat a_{rk}\frac{e^{\lambda_r\Delta T_d}-e^{\lambda_r\Delta T_d(\nu_d + 1)}}{1-e^{\lambda_r\Delta T_d}}}_{=:\ R_k}\biggr]\\
	&& = \nu_d \sum_{k=0}^2 \hat a_{0k}\left(1-e^{-q_k S}\right) + \sum_{k=0}^2 R_k\left(1-e^{-q_k S}\right)
\end{eqnarray*}
And analogously:
\begin{eqnarray*}
	(*) && = \nu_d\cdot \hat a_{03} + \sum_{r=1}^3 a_{0r}\hat a_{r3} \frac{e^{\lambda_r\Delta T_d}-e^{\lambda_r\Delta T_d(\nu_d + 1)}}{1-e^{\lambda_r\Delta T_d}}\\
	&& = \nu_d\cdot \hat a_{03} + R_3
\end{eqnarray*}
So we obtain:
\begin{equation*}
	\sum_{n=1}^{\nu_d} \text{P}(C_n) \leq \nu_d\biggl[\hat a_{03} + \sum_{k=0}^2 \hat a_{0k}\left(1-e^{-q_k S}\right)\biggr] + R_3+\sum_{k=0}^2 R_k\left(1-e^{-q_k S}\right) =: \mathfrak{p}_d^O
\end{equation*}

Thus, the probability to be contacted by a planet around an O-type star at distance $d$ lightyears can be estimated by an upper bound $\mathfrak{p}_d^O$. Repeating these calculations analogously for the remaining star types, we obtain upper bounds $\mathfrak{p}_d^B,...,\mathfrak{p}_d^M$, and the total probability to be contacted by any planet at distance $d$ lightyears can be estimated via
\begin{equation*}
	\sum_{n=1}^{N_d} \text{P}(C_n) \leq \mathfrak{p}_d^O + ... + \mathfrak{p}_d^M =: \mathfrak{p}_d
\end{equation*}
Finally, by summing over all distances within the Detection Radius, we obtain an upper bound on the Contact Probability $p$, the \textbf{Contact Probability Bound} $\mathfrak{p}$:
\begin{equation}\label{Eqn:p_bound}
	p \leq \sum_{d=1}^D \sum_{n=1}^{N_d} \text{P}(C_n) \leq \sum_{d=1}^D \mathfrak{p}_d =: \mathfrak{p}
\end{equation}

\printbibliography

\end{document}